\documentclass[prd,twocolumn,showpacs,nofootinbib]{revtex4}
\usepackage{times}
\usepackage{natbib}
\usepackage{epsfig}
\usepackage{amsmath}
\usepackage[a4paper,left=1.2cm,right=1.2cm]{geometry}

\newcommand{\beeq}{\begin{equation}}
\newcommand{\eneq}{\end{equation}}
\newcommand{\bear}{\begin{eqnarray}}
\newcommand{\enar}{\end{eqnarray}}
\newcommand{\up}[1]{{\rm #1}}
\newcommand{\gbar}{\bar g}

\newcommand{\dobs}{\delta_\up{obs}}

\newcommand{\dz}{\delta z}

\newcommand{\ddL}{\delta\mathcal{D}_L}
\newcommand{\apjl}{Astrophys. J. Lett.}
\newcommand{\aj}{Astron. J.}
\newcommand{\aap}{Astron. Astrophys.}
\newcommand{\mnras}{Mon. Not. R. Astron. Soc.}

\newcommand{\phrd}{Phys. Rev. D}

\newcommand{\natj}{Nature}

\begin{document}

%Title
\title{Connection between Newtonian simulations and general relativity}

%Authors
\author{Nora Elisa Chisari$^1$}
\altaffiliation{nchisari@astro.princeton.edu} 
\author{Matias Zaldarriaga$^2$}
\affiliation{$^1$ Department of Astrophysical Sciences, Princeton University, 4 Ivy Lane,
Princeton, New Jersey 08544, USA}
\affiliation{$^2$ School of Natural Sciences, Institute for Advanced Study,
Olden Lane, Princeton, New Jersey 08540, USA}

%Abstract
\begin{abstract}
On large-scales, comparable to the horizon, the observable clustering properties of galaxies are affected by various general relativistic effects. To calculate these effects one needs to consistently solve for the metric, densities and velocities in a specific coordinate system or gauge. The method of choice for simulating large-scale structure is numerical N-body simulations which are performed in the Newtonian limit. Even though one might worry that the use of the Newtonian approximation would make it impossible to use these simulations to compute properties on very large-scales, we show that the simulations are still solving the dynamics correctly even for long modes and we give formulas to obtain the position of particles in the conformal Newtonian gauge given the positions computed in the simulation. We also give formulas to convert from the output coordinates of N-body simulations to the observable coordinates of the particles.
\end{abstract}

%PACS: cosmology,relativistic astrophysics,redshift,clusters galaxy

\maketitle

%Accepted May 6 2011

\section{Introduction}%****************************************************************

The study of the fluctuations in the distribution of matter in the Universe and its evolution through cosmic history has become one of the major tools in cosmology. The properties and time evolution of the large-scale structure depend on the cosmological parameters and on the initial conditions for the hot big bang. Many of the parameters of the currently favored cosmological model have been determined by matching the observed properties of the distribution of mass through cosmic history with the model calculations. 

Galaxies serve as tracers of the underlying matter distribution. Significant efforts have been made to understand their connection \citep[][]{Kaiser84,Rees85,Bardeen86,Mo96} and to generate estimates for cosmological parameters from
the recovered matter power spectrum \citep[][]{Reid10}. In the last decades, redshift surveys such as the Sloan Digital Sky Survey (SDSS) \citep[][]{York00} and the Two-degree Field Galaxy Redshift Survey (2dFGRS) \citep[][]{Colless01} have resulted in detailed maps of the large-scale distribution of galaxies across very large volumes. The future promises even larger surveys as a result of efforts to improve measures of the so-called baryon acoustic oscillation signal in the clustering of matter to further constrain the properties of the dark energy \citep[][]{Eisenstein05,Percival10}. Surveys are beginning to probe scales comparable to the horizon at the redshift of the galaxies being observed. 
 
Until recently, predictions for observables in galaxy surveys had been done entirely in the Newtonian limit. \citet[][]{Matsubara2000} included gravitational lensing effects on the correlation functions of galaxies and quasars as applied to SDSS. More recently, work by \citet[][]{yoo09,Yoo10} made a more detailed treatment of general relativistic effects. These become relevant as the scales probed by the survey approach the horizon scale. The overdensity in the galaxy distribution $\dobs$ is given by
\bear
\label{eq:dobs}
\dobs&=&b~(\delta_m-3~\dz)+
A+2D+(v^i-B^i)e_i+E_{ij}e^i 
e^j\nonumber  \\
&-&(1+z){\partial\over\partial z}~\dz-2~{1+z\over Hr}~\dz-\dz
-5p~\ddL-2~\kappa
\nonumber \\
&+&{1+z\over H}{dH\over dz}~\dz+2~{\delta r\over r}~,
\enar
where $H$ is the Hubble constant; $\dz$, $\delta r$ and $\ddL$ are the fluctuations in the redshift, distance along the line of sight, and luminosity distance relative to the unperturbed universe; $\kappa$ is the lensing convergence; $p$ gives the slope of the galaxy luminosity function; $b$ is the bias; $e_i$ the direction of propagation of the photon; and $A$, $B_i$, $D$ and $E_{ij}$ are metric components: 
\bear
\label{eq:metric}
ds^2&=&-a^2\left(1+2A\right)d\eta^2-2~a^2B_i ~d\eta ~dx^i \\
&+&a^2\left[(1+2D)\gbar_{ij}+2E_{ij}\right]dx^i dx^j~.
\nonumber
\enar
with $\gbar_{ij}$ the metric tensor for three-space in a homogeneous universe and $\eta=\int dt/a(t)$ is the conformal time in terms of the scale factor.
These formulas exhibit many of the relativistic effects that are common in calculation of the anisotropies in the cosmic microwave background (CMB). For example the observed redshift of a source is given by  
\bear
\label{eq:gSW}
1+z_{obs}&=&\left({a_o\over a_s}\right)\bigg\{
1+\bigg[(v_i-B_i)~e^i-A\bigg]_o^s \\
&-&\int_0^{r_s}dr\left[(A'-D')-(B_{i|j}+E'_{ij})
~e^i e^j\right]\bigg\}~, \nonumber 
\enar
where the prime indicates the derivative with respect to conformal time, the vertical bar is the covariant derivative with respect to $\bar{g}_{ij}$, $r_s=r(z_s)$ is the comoving line of sight distance to the source galaxies at~$z_s$, $v_i e^i$ is the line of sight peculiar velocity, and $a_o$ and $a_s$ are the values of the scale factor at the time of observation and light-emission respectively. The first square bracket represents the redshift-space distortion by peculiar velocities, frame dragging, and gravitational redshift, respectively. The first round bracket in the integral also represents the gravitational redshift, arising from the net difference in gravitational potential due to its time evolution for the duration of photon propagation, and this effect is referred to as the integrated Sachs-Wolfe (ISW) effect in the CMB literature. The last terms in the integral represent the tidal effect from the frame dragging and the ISW effect from the time evolution of the primordial gravity waves. 

The complete set of formulas needed to predict the observed clustering properties of galaxies on very large-scales can be found in \citet[][]{yoo09}. It is clear that the calculation requires consistently solving the general relativity dynamics in a particular coordinate system or gauge. On the other hand N-body simulations are the method of choice to compute predictions for the large-scale distribution of galaxies but these simulations are done in the Newtonian limit. It is appropriate to ask how the output of simulations can be used to compute the different terms in Eq. (\ref{eq:dobs}) and even whether this can be done at all given that the simulations are run using Newtonian dynamics. 

The drive on the observational side to map larger and larger volumes of the Universe and the exponential increase in computer power have also resulted in numerical simulations of ever increasing size. Typical cosmological simulations evolve the particles starting at $z \sim 100$, when the size of the horizon is $\sim 1.5$Gpc. Box sizes vary and can be as large as $\sim 0.5-3$Gpc comoving and the number of particles is of order $\sim 10^9-10^{10}$. Examples of some of the biggest simulations to date are the Millennium Simulation \citep[][]{springel05} run in a box of comoving size $500h^{-1}$Mpc, the Marenostrum Numerical Cosmology Project \citep[][]{gottlober06} and the Hubble Volume project \citep[][]{evrard02} run in a box of $3000 \times 3000 \times 30$ Mpc. Further examples are found in \citet[][]{Colberg00}, \citet[][]{Park05} and recently in \citet[][]{Cai10}. Some of these simulations are started at an initial time when the horizon actually lies within the box. Clearly, we need a way to match cosmological N-body simulations with the general relativistic
variables in Eq. (\ref{eq:dobs}).  

In addition to asking how to use the outputs of numerical simulations to compute the various terms in Eq. (\ref{eq:dobs}) one may wonder if numerical simulations are solving the correct dynamical equations. We might suspect that the Newtonian simulations are working in the so-called conformal Newtonian gauge, in which the line element is given by
\beeq
ds^2=a^2(\eta)[-(1+2\phi_N)d\eta^2+(1-2\phi_N)\delta_{ij}dx^idx^j]
\eneq
in the absence of anisotropic stress and where $\phi_N$ coincides with the Newtonian potential only on small scales. In fact, the analogue of the Poisson equation in the conformal Newtonian gauge reads:
\beeq
\nabla^2\phi_N-3\mathcal{H}(\phi_N'+\mathcal{H}\phi_N)=4\pi G a^2\delta\rho_N
\label{eq:poisson_start}
\eneq
and thus differs from the standard Poisson equation on large-scales. Here, $\mathcal{H}=a'/a$ is the conformal Hubble parameter. 

From Eq. (\ref{eq:poisson_start}), it might appear that simulations are not solving correctly for the gravitational potential for scales comparable to or larger than the horizon. Previous work on this subject has focused on comparing the general relativity equations to the Newtonian equations up to some given order in perturbation theory \citep[][]{Hwang06}. We will show in this work that a more direct approach is possible. 
We will analyze the situation in detail and conclude that N-body simulations are solving for the potential correctly but that the location of the particles needs to be corrected if they are to be interpreted as the particle coordinates in the conformal Newtonian gauge. Finally we will give formulas to recover observable coordinates directly in terms of the output of N-body simulations.

%%%%%%%%%

\section{Evolution equations}%***************************************************
\label{einstein}

As structure in the Universe develops the density contrast becomes larger and larger exceeding unity at the so-called nonlinear scale. 
Properly modeling this process on small scales, of order the nonlinear scale or smaller requires numerical simulations. However, because the primordial curvature fluctuations, the seeds for structure formation, are so small, the nonlinear scale is significantly smaller than the horizon. As a result perturbations in the space time remain very small, of order $10^{-5}$ or smaller\footnote{On sufficiently small scales baryons can collapse to form relativistic objects such as neutron stars or black holes around which the space time metric fluctuations are large. This has negligible effects on the length scales considered in this paper}. 

Thus, to study structure formation we need only consider small perturbations to the Friedmann-Robertson-Walker metric and we can stay at linear order on those perturbations. In this paper we choose to work in the conformal Newtonian gauge with the line element given by:
\beeq
ds^2=a^2(\eta)[-(1+2\psi_N)d\eta^2+(1-2\phi_N)\delta_{ij}dx^idx^j]
\eneq
where $\psi_N$ represents the Newtonian potential and $\phi_N$, the Newtonian curvature. 

We stress that we are assuming that metric perturbations are small but we are not treating the density perturbations using perturbation theory. 
The structure formation process also results in peculiar velocities for the particles. Because the nonlinear scale is well inside the horizon, these peculiar velocities are small, much smaller than the speed of light. In fact at the nonlinear scale peculiar velocities are of order the Hubble velocity for points separated by a distance of order the nonlinear scale. As a result the kinetic energies of particles do not source gravity in an appreciable way. 

Note that we are considering perturbations around the FRW metric so at lowest order the source for the gravitational potentials in $\delta\rho$ as opposed to the full $\rho$. The kinetic energy corrections are of order $\rho v^2$. It is still true that $\rho v^2 \ll \delta\rho$ on every scale. Thus it is safe to ignore the peculiar motions as a source of gravity. These terms are of course also neglected in numerical simulations run using the Newtonian approximation, but this is a negligible source of error. Including these terms is necessary if one wants to study the back-reaction of cosmological perturbations on to the expansion of the Universe \citep[][]{Baumann10}, but we are not interested in this problem here. 

In the standard Newtonian approximation terms of order $\rho \phi$ are also dropped as sources of gravity. This requires a little bit more thought in our case. Again at lowest order the source of gravity is $\delta \rho$ but this is no longer much larger than $\rho \phi$ on sufficiently large-scales. Thus we need to keep this term. However it is only $\bar\rho \phi$ that needs to be kept as of course $\delta\rho \gg \delta\rho \phi$ on all scales. We will now summarize the evolution equations under these approximations. 

\subsection{Einstein equations}

In the conformal Newtonian gauge the Einstein equations $G_{\mu \nu}=8\pi G T_{\mu \nu}$ are reduced to: 
%Einstein equations *************************************************************************
\beeq
\nabla^2\phi_N-3\mathcal{H}(\phi_N'+\mathcal{H}\psi_N)=-4\pi G a^2(T^0_0-\bar{T}^0_0),
\label{poissonrel}
\eneq
\beeq
[\phi_N'+\mathcal{H}\psi_N]_{,i}=4\pi G a^2 T^0_i,
\label{i0}
\eneq
\bear
\phi_N''+\mathcal{H}(2\phi_N'+\psi_N')& + & (\mathcal{H}^2+2\mathcal{H}')\psi_N\\
-\frac{2}{3}\nabla^2(\phi_N-\psi_N)  & = & \frac{4\pi G a^2}{3}(T^i_i-\bar{T}^i_i),
\nonumber
\label{evolphiGR}
\enar
\bear
\label{aniso}
\partial_i\partial_j[(\phi_N-\psi_N)_{,ij} & - & \frac{1}{3}\delta_{ij}\nabla^2(\phi_N-\psi_N)] =\\
  &  & 8\pi G a^2\partial_i\partial_j(T^i_j-\frac{1}{3}\delta^i_jT^k_k)  .
\nonumber
\enar
%*****************************************************************************************
where $_{,i}$ indicates derivatives with respect to the $i$ coordinate and the background Friedmann equation for a flat Universe with cosmological constant $\Lambda$ has been subtracted,
\beeq
\begin{aligned}
\frac{3\mathcal{H}^2}{2a^2}= - 4\pi G \bar{T}^0_0+ \frac{\Lambda}{2},\\
\mathcal{H}^2+2\mathcal{H}'=-\frac{8\pi G a^2}{3} \bar{T}^i_i.
\label{bkg}
\end{aligned}
\eneq

\subsection{Application to nonrelativistic particles}

As we mentioned when studying structure formation we are primarily interested in nonrelativistic matter. Equation (\ref{aniso}) for $i \neq j$ implies that the anisotropic stress is of order $\rho v^2$ and thus negligible in our approximation. As a result $\phi_N=\psi_N$. The gravitational potential satisfies: 
\beeq
\begin{aligned}
\nabla^2\phi_N-3\mathcal{H}(\phi_N'+\mathcal{H}\phi_N)=-4 \pi G a^2 (T^0_0 - \bar T^0_0).
\end{aligned}
\label{00eq}
\eneq
The energy-momentum tensor for a set of cold dark matter particles with mass $m_a$ \cite{weinberg} is given by: 
\beeq
T^{\mu\nu}=(-g)^{-1/2}\sum_a m_a 
\frac{u^{\mu}_au^{\nu}_a}{u^0_a}\delta_D(\vec{x}-\vec{x}_a)
\eneq
where $u_a^{\mu}$ is the comoving velocity of the particles, $dx^{\mu}/d\eta$ and $g$ is the determinant of the metric. At linear order in the metric and to second order in the three-velocities, $v^i$, this is given by $u_a^0=a^{-1}~{(1-\psi_N~+v_a^2/2)}$ and $u_a^i=a^{-1}v_a^i$. The $00$ component of the stress-energy tensor is related to the density of particles and because metric perturbations are small, we can expand in powers of $\phi_N$ and remain at linear order. To order $v_a^2$, the $00$ component is
\beeq
T^0_0=-a^{-3}\sum_a m_a (1+3\phi_N+v_a^2) \delta_D(\vec{x}-\vec{x}_a(\eta)).
\eneq
As discussed we will neglect the $v^2$ term but need to keep the $\phi_N$ as $\phi_N \bar \rho$ is not negligible on large-scales. We obtain
\beeq
T^0_0=-a^{-3} (1+3\phi_N) \sum_a m_a \delta_D(\vec{x}-\vec{x}_a(\eta))
\label{density}
\eneq
where only the $\phi_N \bar\rho$ piece of the term proportional to $\phi_N$ ever makes any difference. 
Replacing in Eq. (\ref{00eq}),
\bear
\label{00withCDMtensor}
\nabla^2\phi_N-3\mathcal{H}(\phi_N'+\mathcal{H}\phi_N) & + & \frac{3}{2}\mathcal{H}^2 = \\ 
4 \pi G a^2 \rho(1+3\phi_N) & + & \frac{\Lambda a^2}{2}
\nonumber
\enar
where $\rho(\vec{x},\eta)=a^{-3}\sum_a m_a\delta_D(\vec{x}-\vec{x}_a(\eta))$ is the density obtained by naively counting particles in cells at each time step. Given the positions of the particles, Eq. (\ref{00withCDMtensor}) can be solved to obtain the Newtonian potential. 

The positions and velocities of the particles are advanced using the geodesic equation,
\beeq
\frac{d^2\vec{x}_a}{d\eta^2}+(\mathcal{H}-3\phi_N')\frac{d\vec{x}_a}{d\eta}=-\nabla\phi_N(\vec{x}_a).
\label{geodesic}
\eneq
Notice that the term $3\phi_N' {d\vec{x}_a}/{d\eta}$ is always negligible. 

%%%%%%%%%%%%%%%%%%

\section{Initial conditions}%*************************************************************
\label{initial}

In addition to the evolution equations we need to find the initial conditions. This can be done at early enough time using linear theory. We define a growth function for the potential such that in the linear regime
\beeq
\phi_N = b_{\phi}(\eta) \phi_N^{in}.
\label{phigrowth}
\eneq
Given that the spatial components of the stress tensor for the dark matter are negligibly small, the gravitational potential in the matter era satisfies
\beeq
\phi_N''+3\mathcal{H}\phi_N'=0.
\label{onlyphi_sim}
\eneq
The solutions to this equation are well-known \citep[][]{mukhanov05}, giving a constant and a decaying solution for the potential in the linear regime,
which in terms of Eq. (\ref{phigrowth}) is
\beeq
b_{\phi}=C_1+\frac{C_2}{\eta^5},
\eneq
where $C_1$ and $C_2$ are constants. The decaying mode is absolutely negligible at the times of interest and without loss of generality we choose $C_1=1$. 

We can replace the right-hand side of the geodesic equation by the solution of the potential in the linear regime,
\beeq
\frac{d^2\vec{x}_a}{d\eta^2}+\mathcal{H}\frac{d\vec{x}_a}{d\eta}=-\nabla\phi_N^{in}.
\label{lin_geodes}
\eneq
In a matter-dominated regime the homogeneous solutions to Eq. (\ref{lin_geodes}) are given by a constant vector and a decaying solution,
\beeq
\vec{x}_{\rm h}=\vec{B_1}\eta^{-1}+\vec{B_2}.
\eneq
For the particular solution, we choose an ansatz $\vec{x}_{\rm p}~=~b_{\delta}(\eta)~\vec{\psi}_1(\vec{x}_{in})$, as it is usually done in the Zel'dovich approximation \cite{zeldovich}. The labeling of $b_{\delta}(\eta)$ as such will become clear by the end of this section. The equation to be solved is,
\beeq
(b_{\delta}''+\mathcal{H}b_{\delta}')\vec{\psi}_1(\vec{x}_{in})=-\nabla\phi_N^{in}.
\eneq
The right-hand side of the last equation is independent of time, which implies that $b_{\delta}''+\mathcal{H}b_{\delta}'=constant$. The actual value of 
the constant is arbitrary, since it can be absorbed in $\vec{\psi}_1(\vec{x}_{in})$. Thus, $b_{\delta} \propto \eta^2$. (Adding a constant to this solution would not modify the subsequent steps of our paper and is only linked to the choice of initial time.) If we equate the factors that depend on the coordinates, $\vec{\psi}_1(\vec{x}_{in})\propto-\nabla\phi_N^{in}$.

To give the complete solution for the position of the particles we separate the constant vector $\vec{B_2}$ in two components, $\vec{B_2}=\vec{x}_{in}+\vec{\delta x}_{in}(\vec{x}_{in})$, where $\vec{x}_{in}$ are the positions of the particles if they started out distributed uniformly in a mesh. We discard the decaying term and give the position of the particles as a function of time in the linear regime,
\beeq
\vec{x}_a=\vec{x}_{in}+\vec{\delta 
x}_{in}(\vec{x}_{in})+b_{\delta}(\eta)\vec{\psi}_1(\vec{x}_{in}).
\label{ansatz}
\eneq
%
%The Poisson Eq.
To find the value of $\vec{\delta x}_{in}(\vec{x}_{in})$, we resort to the Poisson equation (\ref{00withCDMtensor}). We can evolve $\rho(\vec{x},\eta)$ by means of a transformation of coordinates from the initial particle density
\beeq
\rho(\vec{x},\eta)=\frac{\bar{\rho}}{a^3\|\frac{\partial 
\vec{x}}{\partial \vec{x}_{in}}\|},
\eneq
where $\|\frac{\partial \vec{x}}{\partial \vec{x}_{in}}\|$ is the Jacobian of the transformation and $\bar{\rho}$ is the initial uniform background density. The transformation of coordinates is given by Eq.~(\ref{ansatz}), where $\vec{\delta x}$, $b_{\delta}(\eta)$ and $\vec{\psi}_1$ are unknowns. Since the perturbations are initially small, the density evolves as
\beeq
\rho=\frac{\bar{\rho}}{a^3}(1-\nabla \cdot \vec{\delta 
x}_{in}-b_{\delta}(\eta)\nabla \cdot \vec{\psi}_1).
\label{rhoZeld}
\eneq
We can define contributions to the density perturbation, $\delta$, as related to the displacement fields by
\beeq
\delta^{in}=-\nabla \cdot \vec{\delta x}_{in},
\label{deltain}
\eneq
\beeq
\delta^{Z}=-b_{\delta}(\eta)\nabla \cdot \vec{\psi}_1.
\label{deltaZ}
\eneq
The physical meaning of $b_{\delta}(\eta)$ now becomes clear, as it is identified with the growth function of density perturbations. Indeed, as we expect in the matter-dominated regime, we obtained that $b_{\delta}(\eta)\propto a(\eta) \propto \eta^2$. 

To determine $\delta^{in}$, we replace Eqns. (\ref{rhoZeld}), (\ref{deltain}) and (\ref{deltaZ})
in Eq. (\ref{00withCDMtensor}) to obtain:
\beeq
\begin{aligned}
\nabla^2\phi_N-3\mathcal{H}^2\phi_N = \frac{3}{2} \mathcal{H}^2(3\phi_N+\delta^{in}+\delta^Z).
\end{aligned}
\label{00ZelZeta}
\eneq
At the initial time the first term in the left-hand side cancels with the rightmost term in the right-hand side of the previous equation. We can solve for $\delta^{in}$ from the remaining terms,
\beeq
\delta^{in}=-\nabla \cdot \vec{\delta x}_{in}=-5\phi_N^{in}.
\eneq

In terms of their Fourier components, while $\delta^{in}_k\propto \phi_{N,k}^{in}$, the Zel'dovich component dependency is $\delta^{Z} \propto k^2\mathcal{H}^{-2} \phi_{N,k}^{in}$ Well inside the horizon, in the limit $k\eta \gg 1$,  $\delta^{in}$ becomes negligible as compared to $\delta^{Z}$, the Newtonian density perturbation. When $k\eta \ll 1$ then $| \delta^{in}_k | \gg | \delta_k^Z |$ and as anticipated we cannot neglect this term. 

Finally, we can obtain $\vec{\delta x}_{in}$ by inverting Eq. (\ref{deltain}) in Fourier space, 
\beeq
\vec{\delta x}_{in}=-\int\frac{d^3k}{(2\pi)^3}\frac{i\vec{k}}{k^2}\delta^{in}_k 
e^{i\vec{k} \cdot \vec{x}_{in}}.
\label{eq:dxin}
\eneq

In summary, to perform a cosmological simulation we need to evolve the position and velocities of particles using Eq.~(\ref{geodesic}) and compute the potential using Eqs. (\ref{poissonrel}) and (\ref{density}). At the initial time the cold dark matter particles need to be displaced by an amount given in Eq.~(\ref{eq:dxin}).

%%%%%%%%%%%%%%%%%%%%%%%%%

\section{Comparison to Newtonian cosmological simulations}%*******************************************************
\label{compNewt}

Now that we have a consistent set of equations to solve we can compare them to those used in cosmological simulations to determine whether these simulations can be used to study very long wavelength modes or if they require some change. 

Cosmological Newtonian simulations solve for the potential by means of the Poisson equation
\beeq
\nabla^2\phi_{\rm sim}=4\pi G a^2 \frac{\bar{\rho}}{a^3} \delta_{\rm sim}
\eneq
and move particles according to Newton's law expressed in comoving coordinates,
\beeq
\frac{d^2\vec{x}_a}{d\eta^2}+\mathcal{H}\frac{d\vec{x}_a}{d\eta}=-\nabla\phi_{\rm sim}.
\eneq
This evolution equation is identical to the geodesic equation (\ref{geodesic}) if the gravitational potential were computed correctly (as mentioned before the term $\propto 3{\phi_N}^\prime$ is negligible for modes both large and small compared to Hubble). Thus if the gravitational potential is correct the particle positions are updated properly. 

It is important to determine if there are corrections to the gravitational potential that become important on large-scales. 
The density that sources the Poisson equation in simulations is directly computed by counting particles in cells,
\beeq
\rho_{\rm sim} (\vec{x},\eta)=\bar{\rho}\delta_{\rm sim}=a^{-3}\sum_a m_a \delta_D(\vec{x}-\vec{x}_a(\eta)).
\eneq
Even if the particle positions had been computed correctly, this ``simulation density'' differs from the density in the
conformal Newtonian gauge by a factor $(1+3\phi_N)$. 

Finally in standard cosmological simulation the particles are initially displaced making use of the Zel'dovich approximation, which in the Newtonian case takes the form,
\beeq
\vec{x}_{a}=\vec{x}_{in}+b_{\delta}(\eta)\vec{\psi}_1(\vec{x}_{in}).
\label{ZelNewt}
\eneq
This differs from the displacements we calculated in the previous section - it is missing the $\vec{\delta x}_{in}$.   

Thus at first sight it appears that the gravitational potential is not computed using the correct equation, that the density contrast is missing a term and that the initial displacement of the particles is incorrect. We will now show that in fact all these different ``missing terms" cancel each other so that the gravitational potential is computed correctly. As a result, particle positions are also updated correctly. 
 
Let us look at the situation more carefully. For completeness let us also include a cosmological constant and start by restricting ourselves to the linear regime as in any event the effects we are considering are only relevant on very large-scales. The 
relativistic Poisson equation reads
\bear
\label{bspoisson}
\nabla^2 \phi_N-3\mathcal{H}(\phi_N'+\mathcal{H}\phi_N) & = & \\
\frac{3}{2}\mathcal{H}^2(1+\omega)[3(\phi_N-\zeta_{in})+\delta^Z] & & \nonumber
\enar
where we have introduced the equation of state parameter $\omega= \bar p / \bar \rho$ and we have written $-\nabla \cdot \vec{\delta x}_{in}=\delta^{in}\equiv-3\zeta_{in}$. Notice that $3/2\mathcal{H}^2(1+\omega)=4\pi G \bar\rho_{dm}$ with $\bar\rho_{dm}$ the mean density for the matter. The terms in brackets on the right-hand side correspond to the density contrast calculated in the Newtonian simulations ($\delta^Z$) and the two missing corrections, the one proportional to $\phi_N$ and the one coming from the missing initial displacements ($\zeta_{in}$). 

It is useful to consider the comoving curvature $\zeta$ defined as
\beeq
\zeta=\frac{2}{3}\frac{\mathcal{H}^{-1}\phi_N'+\phi_N}{1+\omega}+\phi_N.
\eneq
It is well known that this comoving curvature remains constant in time on large-scales. For completeness we spell out the derivation in the Appendix. 
Note that $\zeta_{in}$ defined above is nothing other than the initial value for this variable $\zeta$. In the case of a perfect fluid $\zeta$ is constant on all scales larger than the sound horizon, or $k^2 c_s^2 << {\cal{H}}^2$. In general what plays the role of $c_s^2$ is just the typical velocity dispersion that relates the magnitude of the spatial components of the energy-momentum tensor to the density. In our case, it is the velocity dispersion of the dark matter particles induced by the growth of perturbations and thus $\zeta$ remains constant all the way to the non linear scale. 

The relativistic corrections to the Newtonian Poisson equation, $\nabla^2\phi_N=3/2\mathcal{H}^2(1+\omega)\delta^Z$, are evidenced from subtracting this expression from Eq. (\ref{bspoisson}).
Notice that the difference between the relativistic terms in the right- and left-hand sides of the Poisson equation is nothing other than:
\bear
3\mathcal{H}(\phi_N'+\mathcal{H}\phi_N)+\frac{9}{2}\mathcal{H}^2(1+\omega)(\phi_N-\zeta_{in})=\\
\nonumber
=\frac{9}{2}\mathcal{H}^2(\zeta-\zeta_{in}).
\enar

Thus the additional terms in the relativistic Poisson equation cancel each other for modes larger than the nonlinear scale. Furthermore, because the nonlinear scale is well inside the horizon once the cancellation begins to fail the additional terms are very small, of order the ratio of the nonlinear scale to the horizon squared. Thus the gravitational potential in Newtonian  simulations coincides with the one in the conformal Newtonian gauge even on very large-scales and thus the dark matter particles are being displaced correctly. 

We have showed that Newtonian N-body simulations calculate the correct gravitational potential and displace the particles correctly. The coordinates of the particles however are not the coordinates in the conformal Newtonian gauge as they are missing the initial displacement. Thus the dictionary between conformal Newtonian gauge variables and numerical simulations is:
\begin{eqnarray}
\phi_N&=&\phi_{\rm sim}, \\
\vec{v}_N &=& \vec{v}_{\rm sim}, \\
\vec{x}_N &=& \vec{x}_{\rm sim} + \vec{\delta x}_{in}.
\label{eq:changeofvar}
\end{eqnarray}

Notice that the reason why the scheme worked was that the ``sound horizon" scale, the scale out to which $\zeta$ was constant in time, was well inside the horizon. This scale is nothing other than the scale that dark matter particles can move since the time of the Big Bang as a result of the peculiar velocities they have. Because the dark matter particles are non-relativistic this distance is well inside the horizon and the mistakes are negligible. Of course a simulation based on Newtonian physics could not work if particles are moving at an appreciable fraction of the speed of light. In general if there is an additional relativistic component the Newtonian simulations would not be computing things properly. The cosmological constant did not cause any problem because even though it is in some sense relativistic it is homogeneous so it does not contribute to the perturbation of the stress tensor. Thus as long as one is modeling nonrelativistic components or a relativistic component that does not cluster one is safe using the Newtonian approximation. Such a Newtonian approximation would not work, for example, during the radiation era where the effective sound speed of the dominant component of the energy density is very close to the speed of light. 

%%%%%%%%%%%%%%%%%%%%%

\section{The comoving gauge}%*************************************************
\label{gaugecomp}

For completeness we will now show that Eq. (\ref{00eq}) can be written making apparent gauge transformations between
the comoving and the conformal Newtonian gauge. 

In Eq. (\ref{bspoisson}), the factor $\frac{\bar{\rho}}{a^3}\delta^Z$ is the density perturbation in Newtonian simulations. The expression in brackets is the conformal Newtonian gauge density perturbation. Indeed rearranging the terms we have
\bear
\nabla^2\phi_N-3\mathcal{H}(\phi_N'+\mathcal{H}\phi_N) & = & \\
4\pi G a^2[\frac{\bar{\rho}}{a^3} \delta^Z+3(\phi_N-\zeta_{in})\rho (1+\omega)]. \nonumber
\enar
As long as $\zeta$ is constant, we can recognize in the right-hand side the gauge transformation for the density 
perturbation between the comoving and the conformal Newtonian gauge,
\beeq
\delta\rho_N= \delta 
\rho_C+3(\phi_N-\zeta)\rho (1+\omega).
\label{transf_CCN}
\eneq
This is a well-known relation \citep[][]{hu04} and in this context it implies that the density
perturbation that Newtonian simulations are obtaining is the one in the comoving gauge.

%%%%%%%%%%%%%%%%%%%%%%%

\section{Observable coordinates}%*******************************************************************

In an inhomogeneous universe, the observed positions of the particles in the simulation are modified due to effects such as the Sachs-Wolfe effect, gravitational lensing, and peculiar velocities. The net result is that photons from a source follow a path that is perturbed with respect to the light cone of an observer in a homogeneous universe. Consider a comoving observer in an inhomogeneous universe with a velocity given by $u^{\mu}=((1-\psi_N)/a,v^{i}/a)$. The direction of observation is $\hat{\bf n}$, defined by $(\theta,\varphi)$ in spherical coordinates, but due to the perturbations to the photon path, the direction toward the point reached by the photon geodesic is actually $\hat{\bf s}$, corresponding to $(\theta+\delta\theta,\varphi + \delta \varphi)$. We now take a particle with coordinates $\vec{x}_a(\eta)$ (already taking into account the correction $\vec{\delta x}_{in}$) and we want to know where it crosses the path of the photons going toward the observer.

Our aim in this section is to correct the positions of the particles in the simulation according to the perturbations in the light cone and to obtain their ``observable'' coordinates. A given particle will be observed when it intersects the light cone of the observer at a certain $\tilde{\eta}$. The unperturbed photon path in parametrized by $r(\eta)=\eta_0-\eta$ and constant angular coordinates that coincide with $\theta_a(\tilde{\eta})$, $\varphi_a(\tilde{\eta})$. The intersection will occur when \citep[][]{Matsubara2000,Yoo10}
\beeq
r_a(\tilde{\eta})=\eta_0-\tilde{\eta}+2\int_{\eta_0}^{\tilde{\eta}}\phi_N d\eta,
\label{eq:etatilde}
\eneq
\beeq
\theta_a(\tilde{\eta})=\theta + \delta \theta=\theta-2\int_0^{r(\tilde{\eta})}d\chi\frac{r(\tilde{\eta})-\chi}{r(\tilde{\eta})\chi}
\frac{\partial \phi_N}{\partial \theta},
\eneq
\beeq
\varphi_a(\tilde{\eta})=\varphi+\delta \varphi=\varphi-2\int_0^{r(\tilde{\eta})}d\chi\frac{r(\tilde{\eta})-\chi}{r(\tilde{\eta})\chi \sin^2\theta_a(\tilde{\eta})}\frac{\partial \phi_N}{\partial \varphi},
\eneq
where $\eta_0$ is the conformal time at the origin and the integrals are taken along the unperturbed light cone (Born approximation). 

The observed redshift of the particle is also different from the one
that we would measure in a homogeneous universe. The transformation
is given by conservation of energy, Eq. (\ref{eq:gSW}). This allows us to write the
set of observable coordinates of the particles as
\bear
z_{obs}=\frac{a(\eta_0)}{a(\tilde{\eta})}[1+V(s)&-&V(0)-\phi_N(s)+\phi_N(0)] \\ 
\nonumber
&&- 2\int_0^{r(\tilde{\eta})}d\chi\frac{d\phi_N}{d\eta}-1 ,\\
\theta_{obs} &=& \theta_a(\tilde{\eta}) -\delta\theta ,\\
\varphi_{obs} &=& \varphi_a(\tilde{\eta})- \delta\varphi .
\label{eq:changetoobs}
\enar
where $V$ indicates the peculiar velocity projected on $\hat{n}$. 
The terms $\phi_N(0)$ and $V(0)$ produced by the gravitational potential and velocity of the observer contribute to the monopole and dipole anisotropies making these terms in practice not useful as cosmological probes.

%%%%%%%%%%%%%%%%%%%%%%%%%%%%%%%%%

\section{Summary}%******************************************************************
\label{conclusions}

We have given a dictionary for how to use the outputs of numerical simulations run using Newtonian dynamics to 
compute the clustering properties of matter even on scales comparable to the horizon. We have shown that as long as there is a large separation between the length scale at which the comoving curvature $\zeta$ starts to evolve with time and the scale of the horizon, the output of calculations based on Newtonian dynamics can be used even on very large-scales provided one reinterprets the coordinates of the particles. 
In standard large-scale structure simulations this separation of scales results from the fact that the nonlinear scale is well inside the horizon, but in general it will occur if all species that cluster are non-relativistic and the density perturbations are small. We gave formulas to compute the coordinates of particles in the conformal Newtonian gauge given the outputs of a simulation and to correct their positions to observable coordinates from the same outputs. 

%*****************************************************************************************************************
\acknowledgments
We acknowledge useful discussions with Jaiyul Yoo, A. Liam Fitzpatrick and J. Richard Gott III. 
M.~Z. is supported by the National Science Foundation under PHY-0855425, AST-0506556 and AST-0907969,
and by the David and Lucile Packard Foundation and the John D. and Catherine T.~MacArthur Foundation.

\vskip 2cm

\vfill
\appendix*
\section{Conservation of the comoving curvature}

Consider a cosmological fluid with energy-momentum tensor
\beeq
T^{\alpha}_{\beta}=(p+\rho)u^{\alpha}u_{\beta}+p\delta^{\alpha}_{\beta}
\eneq
a given equation of state,
$p(\rho)$, and speed of sound, $c_s^2=~\frac{dp}{d\rho}$. The evolution
equation for the potential is given by replacing in
Eq. (\ref{evolphiGR}). 
\beeq
\phi_N''+3\mathcal{H}\phi_N'+(2\mathcal{H}'+\mathcal{H}^2)\phi_N=4\pi G a^2 \delta p,
\eneq
where $\delta p \delta^i_j=-\delta T^i_j$ are the pressure fluctuations. For 
adiabatic perturbations, $\delta p = c_s^2 \delta \rho$, then
\bear
\nonumber
\phi_N''+3\mathcal{H}(1+c_s^2)\phi_N'-c_s^2\nabla^2\phi_N\\
+[2\mathcal{H}'+(1+3c_s^2)\mathcal{H}^2]\phi_N=0.
\label{onlyphi}
\enar
In the previous equation, $c_s/\mathcal{H}$ is the size of the sound horizon.
Consequently, the term of order $c_s^2k^2$, when compared to terms of order 
$\sim \mathcal{H}^2 $ or $\sim c_s^2 \mathcal{H}^2$, is only relevant
when the typical size of the perturbation is smaller than the sound
horizon.

Long wavelength solutions to Eq. (\ref{onlyphi}),
characterized by $kc_s\eta \ll 1$, are easier to address in terms of
a conserved quantity, the comoving curvature $\zeta$,
given by \citep[][]{mukhanov05}
\beeq
\zeta=\frac{2}{3}\frac{\mathcal{H}^{-1}\phi_N'+\phi_N}{1+\omega}+\phi_N.
\label{defcomcurv}
\eneq
Following \citep[][]{mukhanov05}, to prove that the comoving curvature is conserved
we define a new variable
\beeq
u\equiv\exp\left[\frac{3}{2} \int (1+c_s^2)\mathcal{H}d\eta\right]\phi_N.
\eneq
The evolution equation obtained for $u$ from Eq. (\ref{onlyphi}) is then
\beeq
u''-c_s^2\nabla^2u-\frac{\theta''}{\theta}u=0,
\eneq
%this is the general evolution eqn, not only long wavelengths.
where $\theta\equiv\frac{1}{a}[\frac{2}{3}(1-\frac{\mathcal{H}'}{\mathcal{H}^2})]^{-1/2}$.
In the case of long wavelength perturbations, the solution
to the evolution equation is 
\beeq
u = A_1 \theta + A_2 \theta \int \frac{d\eta}{\theta^2}
\eneq 
where $A_1$ and $A_2$ are constants.
It can be shown that
\beeq
\zeta=\frac{2}{3}\sqrt{\frac{3}{8\pi G}}\theta^2\left(\frac{u}{\theta}\right)'
\eneq
reduces to the same expression as Eq. (\ref{defcomcurv}) and remains constant
{\it outside of the sound horizon}.

\end{document}